\documentclass[sigconf,9pt,screen,balance=true]{acmart}

\usepackage{layouts}
\newcommand{\measurept}[1]{\the#1}

\usepackage[british]{babel}
\usepackage{siunitx}
\usepackage{enumitem}
\usepackage{cleveref}

\usepackage{tabularx}
\usepackage{booktabs}


\title{Energy-Proportional Serverless Computing on SBC Clusters}

\author{Yuliya Sasinskaya}
\authornote{Both authors contributed equally to this research.}
\orcid{0009-0006-3003-5732}
\affiliation{
  \institution{Technische Universit\"at Berlin}
  \city{Berlin}
  \country{Germany}
}
\email{yus@3s.tu-berlin.de}

\author{Natalie Carl}
\authornotemark[1]
\orcid{0009-0000-5991-9255}
\affiliation{
  \institution{Technische Universit\"at Berlin}
  \city{Berlin}
  \country{Germany}
}
\email{nc@3s.tu-berlin.de}

\author{Niklas Kowallik}
\orcid{0009-0006-9839-1864}
\affiliation{
  \institution{Technische Universit\"at Berlin}
  \city{Berlin}
  \country{Germany}
}
\email{nk@3s.tu-berlin.de}

\author{David Bermbach}
\orcid{}
\affiliation{
  \institution{Technische Universit\"at Berlin}
  \city{Berlin}
  \country{Germany}
}
\email{db@3s.tu-berlin.de}

\begin{abstract}
    Serverless computing promises to run code only on demand, yet the servers beneath it draw most of their power while idle, so energy consumption does not scale to zero with the workload.
    We show that the statelessness of serverless functions makes it possible to extend scale-to-zero from function instances to physical nodes, enabling energy proportionality.
    Our platform, built on tinyFaaS for clusters of single-board computers, treats node power state as part of worker selection, powering workers off when idle and back on under load.
    On a physical cluster, our approach reduces mean power draw by 39.6\,\% on a bursty, trace-based workload and by 46.7\,\% on a sparse one, without a detectable change in median latency on the bursty workload.
    The cost lands in the tail, where invocations that arrive while all workers are off wait about a minute for a node to boot.
    Under continuous load, the platform behaves like an always-on cluster.
\end{abstract}


\copyrightyear{2026}
\acmYear{2026}
\setcopyright{cc}
\setcctype{by-nc-nd}
\acmConference[Mid4CC '26]{4th International Workshop on Middleware for the Computing Continuum}{December 14--18, 2026}{Tarragona, Spain}
\acmBooktitle{4th International Workshop on Middleware for the Computing Continuum (Mid4CC '26), December 14--18, 2026, Tarragona, Spain}
\acmDOI{10.1145/3861040.3862557}
\acmISBN{979-8-4007-3105-1/2026/12}

\begin{document}
    \maketitle

    \section{Introduction}\label{sec:introduction}

Serverless computing is a cloud computing model where developers deploy applications as small, stateless functions that the cloud runs only when an event calls for them.\!\footnote{
    Strictly speaking, serverless computing also encompasses other managed backend services; for readability, we use the term to refer to Function-as-a-Service only.
}
The platform takes care of everything else, from provisioning and scheduling to autoscaling and isolation~\cite{jonas2019berkeley}.
Arguably, scale-to-zero is among the model's most compelling properties: when there is no work, no function instances run.
The workloads platforms must serve are, in turn, far more granular and bursty than the long-running services that characterise previous cloud programming models.
How efficiently platforms handle this matters more than ever, as the electricity demand of the data centres they run on is surging, with growing consequences for carbon emissions and grid capacity~\cite{iea2025energy}.
The International Energy Agency projects this demand to more than double, from \qty{415}{\tera\watt\hour} in 2024 to \qty{945}{\tera\watt\hour} by 2030~\cite{iea2025energy}.
In principle, serverless computing should be ideally placed here.
A model that runs code only on demand ought to consume energy only on demand, too.
Whether it does comes down to energy proportionality, power consumption that scales with useful work~\cite{barroso2007energy}.

In practice, however, it does not.
Invocations are short and highly bursty, so platforms overprovision capacity and keep instances warm to meet latency objectives~\cite{alquraan2026drops,shahrad2020serverless,nasiri2026ibm}.
Even allocated resources sit largely idle.
In production traces from Huawei's public cloud, roughly two thirds of requests leave more than half of their CPU allocation unused, and about three quarters do the same with memory~\cite{lin2026demystifying,joosen2025cold}.
As a result, servers spend most of their time well below peak utilisation, where they are least efficient, or idle between requests~\cite{barroso2007energy}.
At such low utilisation, the fixed, load-independent share of server power dominates.
Across SPECpower submissions since 2023, even modern servers draw a median \qty{22.5}{\percent} of their peak power when idle, and at \qty{10}{\percent} load, this idle baseline still makes up \qty{64}{\percent} of their power draw~\cite{specpower2026data}.
A function's energy footprint is thus shaped less by the work performed during invocations than by the baseline power drawn in anticipation of future ones.
This runs counter to the promise of scale-to-zero; function instances scale to zero, but energy consumption does not.

Existing power management acts within the server: idle and performance states, DVFS, and power gating reduce dynamic power but leave a substantial static floor for as long as the server remains on~\cite{kaxiras2008power,barroso2026datacenter,orgerie2014survey}.
Switching servers off entirely would remove this floor, but is generally avoided.
Servers hold state that must be migrated, and waking them takes too long to absorb load spikes~\cite{barroso2007energy}.
Moreover, these mechanisms react to local signals, decoupled from the serverless platform, which has the best view of when capacity will be needed.
On IBM's serverless platform, for example, nearly \qty{60}{\percent} of workloads keep at least one instance permanently warm~\cite{nasiri2026ibm}, pinning memory that the hardware cannot know is idle.

Serverless computing removes the first obstacle and exposes the information needed to manage the second.
Because functions are stateless, a node can be drained and powered down without migrating state, and because the platform already decides where functions run, it can equally decide which nodes are powered.
Only granularity remains; aggregate static power changes only in steps of one node's static power.
A cluster of many small, independently switchable nodes, such as single-board computers (SBCs), makes these steps fine enough for cluster power to follow demand, extending scale-to-zero from function instances to physical nodes, down to the residual floor of the network fabric and control plane.
Powering down whole nodes is not new in itself~\cite{meisner2009powernap,krioukov2010napsac,chase2001energy}; what serverless changes is that statelessness, the property defining its workloads, is precisely the one node-level switching requires.

In this paper, we ask: to what extent can platform-controlled node-level power switching make a serverless cluster's power consumption proportional to offered load, and at what cost in latency?
To answer this, we present a serverless platform for SBC clusters that treats per-node power state as a first-class scheduling decision.
The central challenge is that switching is not free.
Nodes take seconds to boot and draw power while doing so, so the platform must keep enough capacity powered to meet latency objectives as load changes.
We make the following contributions.
\begin{itemize}[leftmargin=*]
    \item We argue that statelessness lets scale-to-zero extend from function instances to physical nodes (\cref{sec:background}).
    \item We design a platform for SBC clusters whose worker selection also decides which nodes are powered (\cref{sec:design}).
    \item We evaluate it on a physical SBC cluster, where it draws \qtyrange{39.6}{46.7}{\percent} less power under bursty and sparse load (\cref{sec:evaluation}).
\end{itemize}

    \section{Background}\label{sec:background}

Two questions determine whether node-level power switching can make serverless platforms energy-proportional: which power management mechanisms the underlying hardware already provides, and whether serverless load varies in a way that makes powering off entire nodes feasible.

\subsection{Server power management}

Servers provide power management mechanisms at two levels.
At the component level, idle and performance states, DVFS, and clock and power gating reduce the power drawn by processors, memory, and interconnects when they are underutilised~\cite{kaxiras2008power,barroso2026datacenter}.
Three issues prevent them from reaching their potential with serverless workloads.
First, their main target is the CPU, while power supply losses, fans, chipset, network interface, storage, and memory refresh keep drawing power for as long as the server is powered on.
Second, deeper idle states save more power but take longer to exit; with sub-second, bursty arrivals, idle periods are too short for them to pay off, so hardware mostly remains in shallow states~\cite{meisner2009powernap,barroso2007energy}.
Third, idle does not mean unallocated: DRAM can enter low-power modes when idle~\cite{lee2021greendimm}, but memory holding the state of warm but inactive instances must be retained and refreshed; only the platform knows that this state could be discarded~\cite{shahrad2020serverless,nasiri2026ibm,fuerst2021faascache}.

At the system level, firmware interfaces such as ACPI define sleep and soft-off states, and wake-on-LAN can bring a powered-down server back~\cite{nedevschi2009skilled}.
The capability to switch a server off thus exists, yet operators tend to avoid it, as servers hold state and waking them is slow~\cite{barroso2007energy,lin2013rightsizing}.
What is missing is a decision-maker with the right information.
Hardware and OS governors observe only local, past utilisation~\cite{linux2026cpuidle} and cannot tell whether an idle instance’s memory is still needed, nor move work elsewhere to free a node.
The serverless platform can: it knows which instances are idle and when their keep-alive expires, which functions have latency objectives, how aggregate load develops, and where functions could be placed instead.
The responsibility for managing node power state therefore belongs to the platform, while component-level mechanisms remain useful within the nodes it keeps powered.

\subsection{Short bursts, long cycles}\label{sec:background-load}

For the functions that receive the bulk of the traffic, most inter-arrival times lie below one second~\cite{shahrad2020serverless,nasiri2026ibm}.
A node that takes seconds to boot cannot follow fluctuations at this scale.
Across many functions, however, these bursts largely average out, while daily and weekly patterns persist, as they affect most applications at the same time~\cite{shahrad2020serverless,joosen2023function,nasiri2026ibm}.
Nasiri et al.\ observe this in production data from IBM's serverless platform, where the difference between daily maximum and minimum traffic amounts to about \qty{60}{\percent} of peak traffic on weekdays and \qty{40}{\percent} on weekends~\cite{nasiri2026ibm}.
The load a node-level power manager faces is thus far smoother at the scale of seconds than that of any single function, yet cyclical at the scale of hours, and boot times of less than a minute are small against variation of this kind.

This separates two decisions that operate at different timescales.
At the instance level, platforms predict when the next request for a given function will arrive to decide which instances to keep warm, at timescales from milliseconds to minutes~\cite{shahrad2020serverless,fuerst2021faascache,nasiri2026ibm}.
At the node level, the question is how much hardware must be powered for the aggregate load, which changes over hours.
The two decisions are complementary: instance management determines what runs on the powered nodes, node management determines how many nodes are powered.
Short-term fluctuations are absorbed by \emph{powered headroom}, spare capacity that stays on, so that node switching only has to follow the slow, cyclical component of the load.
Our evaluation deliberately omits this headroom and replays a single bursty function, exposing the full switching cost rather than this smoothed case.

    \section{Design}\label{sec:design}

Our design turns the two conclusions of \cref{sec:background} into mechanisms: deciding where an invocation runs also decides which workers must be powered, and powered headroom can spare invocations the wait for a boot, at the cost of idle power.
We keep the policies governing both deliberately simple and reactive so as to separate the effect of the mechanism from that of any specific prediction strategy, which our platform could incorporate.

\subsection{Architecture and node lifecycle}

Our platform consists of a leader and a set of workers.
The leader runs the control plane.
It receives all invocations, load-balances between workers, and decides which nodes are powered.
Moreover, it is never powered down, as no component would then remain to power the others back on.
The leader is \emph{dispatch-only} and does not execute requests itself.
This keeps the residual floor constant and explicit; a leader that also executed functions would absorb light load indefinitely and introduce additional variance in the size of the residual floor.
Workers, in turn, are the unit of power switching.
As a powered-off worker cannot wake itself, the leader must be able to control each worker's power supply from the outside.
Such capabilities are widely available: servers expose them through baseboard management controllers, for example via IPMI or the more recent Redfish~\cite{intel2015ipmi,dmtf2026redfish}, and switched power distribution units provide them per outlet.
For our SBC test bed, we use simple relays.
As workers hold no state, the leader cuts their power directly, without shutting down the operating system first.

Each worker moves through three states.
(1)~A \emph{sleeping} worker is powered off entirely.
(2)~When the leader wakes it, the worker is \emph{waking}: it is powered on, boots, and reports as healthy, after which the leader deploys all functions to it.
Waking thus amounts to a cold start at node level, and statelessness keeps it cheap in all but time, as the worker needs function code but no state.
(3)~Once deployment succeeds, the worker is \emph{active} and serves invocations until the leader powers it down again.
A worker that fails to wake is excluded from the pool of workers but retried periodically.
Workers hold no functions across power cycles.
Instead, the leader keeps the definition of every function and replicates all of them to every worker it wakes, so that every active worker can serve every function.
We expect the set of deployed functions to remain small in the context of SBC clusters.
Should it outgrow a worker's capacity, however, the leader can replace this replicate-everywhere strategy with explicit decisions about which functions to place on which worker.

\subsection{Power-aware worker selection}

The leader selects a worker for each invocation individually, falling back through three options.
It first chooses the least-loaded active worker with fewer than~$L$ invocations in flight.
If no such worker exists, it wakes a sleeping worker and forwards the invocation once that worker is active.
Only when no worker is left to wake does it fall back to the least-loaded but saturated worker.
Workers are thus woken only under pressure, so the number of powered workers follows load rather than anticipating it.
The invocation that triggers a wake pays the price of this reactive approach, as it waits for the entire boot.
We accept this wait deliberately, as it exposes the full cost of switching.

\subsection{Releasing workers and powered headroom}

A controller on the leader periodically checks every active worker and powers it down once it has served no invocation for an idle timeout~$\tau$ and has none in flight.
Because functions are stateless, powering down requires no migration, as a worker without invocations in flight holds nothing that must be preserved.
Choosing~$\tau$ above the boot time protects against oscillation, as a worker released during a brief lull would otherwise be woken again shortly after, paying the full switching cost for little saving.
Deployed functions stay warm until their worker is powered down, so an active worker never incurs a function-level cold start; node-level power management replaces scale-to-zero for individual function instances.

The controller never powers down workers below a floor of~$H$ active workers.
This floor is the powered headroom described in \cref{sec:background-load}, capacity kept on to absorb short-term fluctuations at the cost of its idle power.
Here, the right value depends on the workload.
As the leader executes no invocations itself, choosing $H = 0$ leaves no powered capacity once load ceases, so the first invocation after an idle period waits for a full boot.
For workloads with long idle periods and latency-tolerant invocations, this maximises savings at an acceptable cost; a larger~$H$ trades idle power for latency.

    \section{Evaluation}\label{sec:evaluation}

To find out what node-level power switching saves and what it costs, we compare our platform against the same cluster held fully powered, under workloads that leave different amounts of idle time.

\subsection{Experiment setup}\label{sec:eval-setup}

We evaluate our platform on a physical cluster of single-board computers, which lets us measure the power of every node directly rather than estimate it.
We will release the prototype, the benchmark client, all raw measurements, and our analysis scripts as open source.

\begin{figure}[t]
    \centering
    \includegraphics[width=\columnwidth]{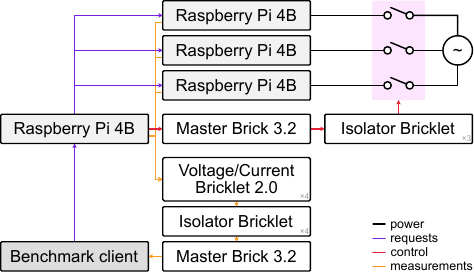}
    \caption{Test bed wiring. A Master Brick on the leader switches each worker's power through a relay; a separate Master Brick on the benchmark client collects voltage and current of all four nodes. Isolator Bricklets separate both paths electrically.}
    \label{fig:wiring}
\end{figure}

\paragraph{Prototype}
We implement our design on tinyFaaS, a lightweight serverless platform for resource-constrained environments, which we extend with cluster support, node-level scale-to-zero, and relay-based power control~\cite{pfandzelter2020tinyfaas}.
We use an idle timeout of $\tau = \qty{60}{\second}$, a load threshold of $L = 10$, no powered headroom ($H = 0$), and a controller interval of \qty{10}{\second}.
Workers run the operating system's default CPU frequency and idle governors, so any savings come in addition to component-level power management.

\paragraph{Test bed}
Our cluster consists of four Raspberry Pi~4B boards, one leader and three workers.
The leader switches each worker's power supply through a relay.
Voltage and current of every node, the leader included, are measured by the benchmark client at \qty{50}{\hertz} (\cref{fig:wiring}).
Control and measurement use separate brick stacks, and the measurements are logged on the benchmark client, so recording power does not load the cluster and power samples share a clock with the request log.

\paragraph{Workloads}
Each invocation runs Sobel edge detection on a \qty{25}{\kilo\byte} JPEG image, as a representative of CPU-bound serverless functions.
We use three workloads that differ in how much idle time they leave.
The \emph{sparse} workload issues one request every two minutes for \qty{20}{\minute}.
The \emph{bursty} workload replays one function from the Azure Functions trace~\cite{shahrad2020serverless}, compressing 240 trace minutes into \qty{30}{\minute}; its 77 invocations arrive in 23 bursts, separated by idle gaps of \qtyrange{67.5}{187.5}{\second} that cover \qty{55}{\percent} of the replay.
We select a function whose idle gaps exceed the idle timeout, as the highest-volume functions in the trace are active every minute.
The \emph{continuous} workload issues \qty{10}{requests\per\second} for \qty{10}{\minute}, leaving no worker idle long enough to be released.

\paragraph{Baseline and runs}
We compare our platform against the same cluster with power switching disabled, which keeps all workers powered.
We run each combination of workload and configuration four times, interleaving configurations to avoid drift, except under continuous load.
We exclude one sparse power-aware run from the energy statistics, as its power trace ends before the run does.
We measure latency at the leader; clients abandon requests waiting longer than \qty{37.5}{\second}, but the leader still records their completion.
We treat each run as one replicate and report \qty{95}{\percent} confidence intervals.
As always-on runs vary far less than power-aware runs, we use Welch's t-interval, which does not assume equal variance, and for relative changes Fieller's method, which also accounts for uncertainty in the always-on reference.

\subsection{Power savings}\label{sec:eval-power}

\begin{figure}[t]
    \centering
    \includegraphics{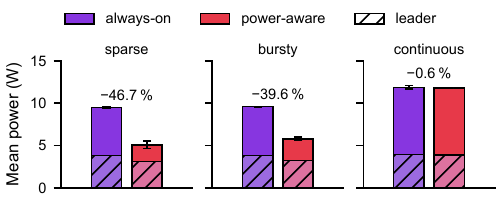}
    \caption{
        Average cluster power per workload.
        The hatched segment shows the leader's share, which varies far less than the workers' share across workloads and configurations.
    }
    \label{fig:mean-power}
\end{figure}

\begin{figure}[t]
    \centering
    \includegraphics{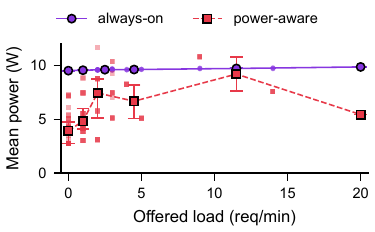}
    \caption{
        Cluster power draw and offered load for the bursty workload.
        Runs are split into consecutive \qty{60}{\second} windows.
        The plot shows the number of requests and the cluster's mean power draw for each window.
        Faint markers are individual windows; lines connect the median per load level (0, 1, 2--3, 4--7, 8--15, and 16--31 requests), with error bars showing the interquartile range.
    }
    \label{fig:power-vs-load}
\end{figure}

\Cref{fig:mean-power} shows mean cluster power per workload.
Under the sparse and bursty workloads, the power-aware configuration draws \qty{46.7}{\percent} (95\,\% CI [42.1, 51.4], $n = 3$) and \qty{39.6}{\percent} (95\,\% CI [37.3, 41.9]) less power than always-on.
Under continuous load, no worker is ever idle for~$\tau$, so all three stay powered after the initial wake, and mean power differs by at most a few percent (\qty{-0.6}{\percent}, 95\,\% CI [$-2.4$, $+1.3$]).

The savings stem from worker time.
Workers are powered \qty{82.9}{\percent} less of the time under the sparse workload and \qty{77.3}{\percent} less under the bursty workload, yet mean power falls by less than half, as the leader never switches off.
The leader's draw rises linearly with the number of powered workers, from \qty{2.80}{\watt} with all workers off by \qty{0.36}{\watt} per worker ($R^2 = 0.99$), consistent with the relays drawing their holding current from the leader's supply; the always-on baseline pays this overhead for all three workers throughout.
The leader and its switching hardware thus form the residual floor that node-level switching cannot remove.

\Cref{fig:power-vs-load} breaks the bursty workload into \qty{60}{\second} windows.
Always-on draws about \qty{9.6}{\watt} regardless of offered load, whereas power-aware rises with it.
Its windows cluster at levels that correspond to zero to three powered workers, and their spread at a given load reflects history, as workers woken by an earlier burst stay powered for up to \qty{60}{\second} after it.
Offered load describes how requests arrive, not how busy the cluster is.
Even the busiest minute, at 20 requests, stays far below the continuous workload, under which about one invocation is in flight on average at 30 times the rate.
The highest load levels rest on only a few windows, each from a single minute of the trace, and hence carry less information than intervals with more data points.

\subsection{Effects on latency}\label{sec:eval-latency}

\begin{figure}[t]
    \centering
    \includegraphics[width=\columnwidth]{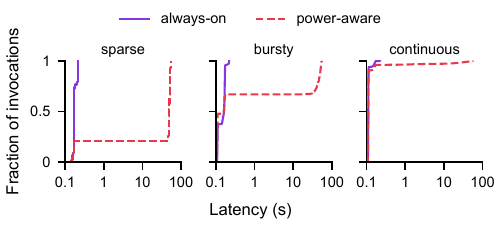}
    \caption{
        Latency per workload, from the leader receiving an invocation to returning its response, including any wait for a wake.
        Distributions cover all successful invocations of all runs.
    }
    \label{fig:latency}
\end{figure}

\Cref{fig:latency} shows the latency distribution per workload.
A worker takes \qtyrange{48}{58}{\second} to wake, with a median of \qty{53}{\second}, and every invocation that arrives while no powered worker has capacity waits for it.
Under the sparse workload, which leaves every worker idle for longer than \qty{60}{\second} between requests, the leader wakes a worker 8 times per run on average, and about four in five invocations wait for a wake.
As a result, median latency rises from \qty{0.17}{\second} to \qty{49.5}{\second}.
This workload is adversarial by construction and marks an upper bound of the latency cost.

Under the bursty workload, the leader wakes a worker 13 times per run, and each wake serves every invocation of the burst that triggered it.
Both distributions coincide for about two thirds of invocations, and we detect no change in the median (\qty{-6.9}{\percent}, 95\,\% CI [$-19.1$, $+5.4$]).
The remaining third arrives at the start of a burst and waits for a wake, which raises the 95\textsuperscript{th} percentile by \qty{51.9}{\second} (95\,\% CI [50.2, 53.6]).

Under continuous load, the power-aware cluster starts with all workers asleep and wakes each of them once at the start of the run.
Requests that arrive during these wakes accumulate on the leader, and some fail with an error when the leader forwards them to a worker that reports ready before it can serve them.
Once the backlog clears after about 56 s, no worker sleeps again and latency matches always-on; over the whole run, the 95th percentile rises by only \qty{11.7}{\percent} (\qty{95}{\percent} CI [9.4, 14.0]).
As the backlog occurs only once, at the start of a run, its share of invocations shrinks as a run gets longer.
Requests waiting for a wake occupy the load generator’s concurrent clients, so we raise its limit from 50 to 800 for the power-aware configuration to keep the backlog from suppressing offered load; always-on never reaches the limit.
At the same limit of 50, power-aware yields the same mean power (\qty{-0.7}{\percent}).

    \section{Discussion}\label{sec:discussion}

Node-level switching makes cluster power follow offered load whenever idle gaps outlast the idle timeout, and it costs latency only where invocations meet a cluster without powered capacity.
Both outcomes are shaped as much by the decisions surrounding the mechanism as by the mechanism itself, namely which worker receives an invocation, how much capacity stays powered, and which hardware must remain on regardless.

\paragraph{Selection policy}
Under continuous load, the power-aware configuration saves almost no power, which reflects the selection policy rather than the switching mechanism.
With about one invocation in flight on average and $L = 10$, a single worker could carry the load, but least-loaded selection spreads invocations so that no worker stays idle for $\tau$.
Packing invocations onto the most loaded worker below $L$ would release the others, at the price of higher contention on that worker and of waits for a wake whenever a burst exceeds $L$.
The policy also ignores invocations already waiting for a wake, so a cluster without capacity can wake several workers at once, as at the start of each continuous run.
In future work, we plan to evaluate the impact of different worker selection policies on overall cluster power consumption.

\paragraph{Hiding the boot}
Invocations that arrive while no powered worker has capacity wait about \qty{53}{\second} for a boot, which accounts for the entire latency cost.
Three extensions could reduce it, at different costs in power.
First, the leader could execute waiting invocations itself while a worker boots, using capacity that the residual floor already pays for.
Second, powered headroom of $H \geq 1$ keeps capacity available at all times, trading the idle draw of each headroom worker for latency.
In our evaluation, we explicitly chose $H=0$ to quantify an upper bound for how much power node-level switching saves.
Third, as daily and weekly cycles dominate aggregate load on public serverless platforms, the predictors they already use to keep instances warm could also be used to wake workers ahead of demand~\cite{fuerst2021faascache,nasiri2026ibm,shahrad2020serverless}.
Our measurements also overstate the latency cost in one respect.
Every power-aware run starts with all workers asleep, so its first invocations wait for a wake that a cluster in steady operation would rarely incur, while always-on starts fully powered.
This start-up wake causes the entire latency cost under continuous load and adds to the tail of the bursty workload, but changes little under the sparse workload, in which most invocations wait regardless.
It also lowers the power-aware configuration's draw during the first minute of each run.

\paragraph{Beyond our test bed}
The residual floor of the leader and its switching hardware is partly avoidable, as our relays draw holding current from the leader's supply, whereas latching relays need power only to switch.
On servers, baseboard management controllers draw standby power while the server is off, and each server is a coarser step, so fine-grained nodes such as SBCs at the edge suit the approach best.
Larger clusters would need a function registry with selective replication instead of replicate-everywhere.
Cutting power without a shutdown loses no application state, but it stresses storage, and one SD card failed during our experiments; read-only root file systems or network boot would avoid this, and the idle timeout also bounds wear.
Finally, AI accelerators are increasingly power-intensive yet often run well below peak; their functions, however, are stateless only nominally, as powering a node down discards model weights whose loading dominates cold starts~\cite{fu2024serverlessllm}.
Whether node-level switching extends to hardware-accelerated serverless computing, for example on SBC-sized edge accelerators, remains open.

    \section{Related work}\label{sec:related}

Adjusting the number of powered servers to load is an established approach to energy proportionality.
Muse concentrates the load of a hosting centre on fewer servers and powers down the rest~\cite{chase2001energy}, NapSAC does the same for web clusters~\cite{krioukov2010napsac}, and Lin et al. derive online algorithms for right-sizing that account for the cost of switching~\cite{lin2013rightsizing}.
PowerNap instead proposes hardware that enters a near-zero-power state within milliseconds, so that a single server can follow idle periods of seconds or less~\cite{meisner2009powernap}.
Cluster autoscalers in public clouds remove idle nodes as well, but release VMs to save cost rather than power down hardware.
These cluster-level approaches depend on statelessness as a property of specific services and on a cluster manager separate from the application, whereas serverless computing makes every function stateless and leaves both worker selection and power control to the same platform.
Our platform therefore needs no hardware support beyond external power control and, with powered headroom, has to follow only the slow component of the load.

Within powered servers, EcoFaaS adapts CPU frequencies to the latency requirements of individual invocations~\cite{stojkovic2024ecofaas}, and GreenWhisk makes the platform itself emission-aware~\cite{serenari2024greenwhisk}.
Other work shifts workloads in time or across regions towards cleaner energy or better utilisation~\cite{wiesner2021lets,gsteiger2024caribou}, although the benefits of such shifting are still debated~\cite{sukprasert2024limitations}.
All of these act on the work or within powered hardware and leave the number of powered nodes unchanged, so they complement node-level switching.

Carl et al. propose executing each invocation on a dedicated SoC that boots for it and shuts down afterwards, removing the need for software isolation as well as idle power~\cite{carl2025towards}.
At the edge, clusters of mobile SoCs already serve as servers~\cite{xu2022renovating,zhang2024more}.
Our platform lies between conventional servers and per-invocation hardware, switching nodes with load rather than per invocation, so that a single boot can serve many invocations.

    \section{Conclusion}
    Our findings indicate that node-level power switching is a viable strategy for reducing the electricity consumption of SBC clusters running serverless workloads.
    Because serverless functions are stateless, the platform can extend scale-to-zero from function instances to physical nodes: on our cluster, this reduces mean power draw by 39.6\,\% on a bursty, trace-based workload and by 46.7\,\% on a sparse one, without a detectable change in median latency on the bursty workload.
    The cost is a boot of about a minute for invocations that find no capacity, which bridging boots at the leader, powered headroom, or prediction could hide.
    Edge sites, which already deploy clusters of small nodes, are a natural place to start.

\begin{acks}
    We thank Sripriya Srikant Adhatarao for her mentorship and many valuable discussions.
    This work was funded in part by the \grantsponsor{BMFTR}{German Federal Ministry of Research, Technology, and Space}{https://www.bmftr.bund.de/} under grant \grantnum{BMFTR}{16IS23068}.
\end{acks}
    \bibliographystyle{ACM-Reference-Format}
    \bibliography{references-shorter}

\end{document}